\documentclass[twocolumn,showpacs,preprintnumbers,amssymb,pra]{revtex4}
\usepackage{graphicx}
\usepackage{dcolumn}
\usepackage{color}
\usepackage{bm}
\begin{document}

\title{Particle motion and gravitational lensing in the metric of a dilaton 
black hole in  a de Sitter universe}
\author{Nupur Mukherjee}
\altaffiliation{Email:nupur@bose.res.in}
\affiliation{S. N. Bose National Centre for Basic Sciences,
Salt Lake, Kolkata 700 098, India}
\author{A. S. Majumdar}
\altaffiliation{Email:archan@bose.res.in}
\affiliation{S. N. Bose National Centre for Basic Sciences,
Salt Lake, Kolkata 700 098, India}

\begin{abstract}
We consider the metric exterior to a charged dilaton black hole in a 
de Sitter universe. We study the motion of a test particle in this metric.
Conserved quantities are identified and the Hamilton-Jacobi method is
employed for the solutions of the equations of motion. At large distances
from the black hole the Hubble expansion of the universe modifies the
effective potential such that bound orbits could exist up to an upper limit
of the angular momentum per mass for the orbiting test particle. 
We then study the phenomenon of strong field 
gravitational lensing by these black holes by extending the standard
formalism of strong lensing to the non-asymptotically flat dilaton-de 
Sitter metric. Expressions for the various lensing quantities are obtained in
terms of the metric coefficients. 
\end{abstract}
\pacs{95.30.Sf, 04.70.Bw, 98.62.Sb}

\maketitle

\section{Introduction}

There has been some renewed interest in the study of 
several different black hole solutions such as the 
Gibbons-Maeda-Garfinkle-Horowitz-Strominger (GMGHS) dilaton black hole 
of string theory\cite{Gibbons}, the
brane world black hole with tidal charge\cite{dadhich}, the Myers-Perry black
hole in the braneworld context\cite{ASM,liddle}, and the 
Einstein-Born-Infeld black hole\cite{borninf}, in recent times.
Such studies are expected to shed light on the character of the strong 
gravitational field and the role of postulated
extra dimensions in string and brane theories. Comparison with
the standard black hole solutions in four dimensions is important.
For example, the braneworld black hole with tidal charge\cite{dadhich} bears
formal resemblance to the Reissner-Nordstrom black hole. The question of
existence of closed orbits is also interesting, and has been recently
investigated\cite{Stojkovic} in the context of braneworld geometry.
In addition, cosmological implications may be significant, as in the case of
the Myers-Perry black hole in
braneworld cosmology due
to its modified evaporation and accretion properties\cite{ASM,liddle}.

The study of black holes in the background of a
cosmological constant, i.e. in the de Sitter space time,
have also attracted some interest for their conjectured relevance in
AdS/CFT correspondence\cite{maldacena}, and due to the phenomenon of
black-hole anti-evaporation \cite{Hawking}.
The dilaton black hole solution in a de Sitter universe has 
been recently derived\cite{CZhang}.
The dilaton of the  GMGHS black hole is a 
scalar field occuring in the low energy limit of the
string theory. It has an important role on the causal 
structure and the thermodynamic properties of the black hole.
Though it was shown earlier that a dilaton black hole solution in the
de Sitter background is not possible with a simple dilaton 
potential\cite{poletti}, Gao and Zhang\cite{CZhang} circumvented this
problem with a dilaton potential being a combination of Liouville type
terms. Such a black hole solution is interesting since it combines stringy
features with the backdrop of the present accelerated expansion of the
universe.

The motivation for the present work is to explore avenues for possible 
signatures
of cosmological expansion on the orbits of massive bodies and light around
dilaton-de Sitter black holes. Gravitational lensing is a widely used tool 
in modern cosmology. Historically,
the theory of gravitational lensing
was primarily developed in the weak field thin-lens approximation
\cite{Sch}, and some recent studies of weak field lensing in braneworld
geometries have also been performed\cite{frolov}. Although these 
approximations are sufficiently accurate to discuss
any present physical observations, the realization
that many galaxies host supermassive
central black holes\cite{Richstone} has spurred the
development of lensing phenomenology in the strong-field 
regime\cite{Fritte}. Bozza et al\cite{Bozzaetal} developed
an analytical technique for obtaining the deflection angle for the 
Schwarzschild black hole. Strong field lensing in the Reisser-Nordstrom 
space-time was then investigated\cite{Eiroaetal}.
Bozza\cite{Bozza} extended the 
analytical theory of strong lensing for a general class of static, 
spherically symmetric, and asymptotically flat metrics and 
showed that the logarithmic
divergence of the deflection angle at the photon sphere is a common
feature of such spacetimes. 

From the viewpoint of future observations, these are exciting times
in the development of gravitational lensing of compact objects.
Newer black hole solutions corresponding to different theoretical models
have appeared in the literature\cite{new}, and  lensing
due to several black hole metrics  have been analysed 
recently\cite{frolov,Fritte,Bozza,Bhadra,Eiroa1,Whisker,NM2,bozza2,Eiroa2}
with the goal of possible observational discrimination in future.
Gravitational lensing by the GMGHS black hole
has been studied by Bhadra\cite{Bhadra}. Subsequently, the strong field
phenomenon of retrolensing has been discussed\cite{Eiroa1}, and the position
and magnifications of relativistic images due to strong lensing by several
braneworld geometries representing the black hole at the centre of our
galaxy have been computed\cite{Whisker,NM2}. Several interesting
features of lensing by rotating black holes have also been 
identified\cite{bozza2}. The theoretical
framework of strong field gravitational lensing is being 
developed further with more suggested applications\cite{petters}.
In order to detect the existence of relativistic images a high
order sensitive instrument with very long baseline interferometry is
required. Planned facilities such as
Maxim can achieve up to $0.1 \mu  arc sec$ resolution in the
X-ray band\cite{Mancini}.  But if future experiments can attain
$0.01 \mu  arc sec$ resolution, and if such 
relativistic images are detected, then lensing observations
will be able to distinguish between the predictions of different black
hole models and probe deeply the character of gravity in the strong field
regime. 

In this paper we present  properties of the motion of particles and light 
in the spacetime of a dilaton-de Sitter black hole. We derive 
equations of motion and 
different conserved quantities for particle motion. We also derive the 
expression for the effective potential which can be
used to obtain bound orbits in this geometry. Analysis of the effective
potential for the dilaton de-Sitter black hole shows that the Hubble expansion
of the universe may significantly contribute to the effective potential 
only at large distances. In presence of the Hubble term, we derive
an upper bound on the value of the angular momentum per mass of the 
orbiting particle that can be in a bound orbit around the black hole. 
We then study strong gravitational
lensing in this spacetime applying the method of Bozza\cite{Bozza} for
the case of a non-asymptotically flat metric for the first time. We 
obtain the expressions for the lensing observables in terms of the 
metric coefficients for this geometry.

The paper is organised as follows. The dilaton-de Sitter black hole, its 
properties,
and various limiting cases are briefly described in Section II.
In Section III we study several aspects of particle
motion in this metric. The effective potential for this metric is obtained
and analysed for the existence of bound orbits in the presence of de Sitter
expansion of the universe. The investigation of strong field 
gravitational lensing is performed 
in the non-asymptotically flat dilaton-de Sitter metric and the 
expressions for the  lensing observables obtained in terms of the metric
coefficients in Section IV.  
We conclude with a summary of our results in Section V.

\section{The dilaton-de Sitter black hole}

The action corresponding to gravity coupled to the dilaton and Maxwell fields
in four dimensions is given by\cite{CZhang}  
\begin{eqnarray}
S=\int d^4x \sqrt{-g}\left[R-2\partial_{\mu}\phi\partial^{\mu}\phi-V(\phi)
               -e^{-2\phi}F_{\mu\nu}F^{\mu\nu}\right]
\end{eqnarray}
where $R$ is the scalar curvature, $F_{\mu\nu}$ is the Maxwell field,
$\phi$ is the dilaton field and $V(\phi)$ is a potential for $\phi$.
Varying the action with respect to metric, Maxwell field and dilation 
field, yields respectively, the Einstein, Maxwell and dilaton field
equations. Though the dilaton coupling to the electromagnetic sector is 
non-zero, the Maxwell equations remain the same as those written in a metric 
with a conformal factor $e^{-\phi}$. Null geodesics are unaffected 
by conformal 
transformations, so that one can continue following null geodesics of the 
original metric in spite of the coupling between photons and dilatons. The 
Maxwell equation can be integrated to give,
$F_{01}=\frac{Qe^{2\phi}}{f^2},$ where $Q$ is the electric charge 

The most general form of any static and spherically symmetric metric is
\begin{eqnarray}
ds^2&=&-U(r)dt^2+\frac{1}{U(r)}dr^2+f(r)^2d\Omega^2.
\end{eqnarray}
Using the above form for the metric, the field equations can be reduced
to three independent equations\cite{CZhang}, i.e.,
\begin{eqnarray}
\frac{1}{f^2}\frac{d}{dr}\left(f^2U\frac{d\phi}{dr}\right)
&=&\frac{1}{4}\frac{dV}{d\phi}+e^{2\phi}\frac{Q^2}{f^4},
\end{eqnarray}
\begin{eqnarray}
\frac{1}{f}\frac{d^2f}{dr^2}&=&-\left(\frac{d\phi}{dr}\right)^2,
\end{eqnarray}
\begin{eqnarray}
\frac{1}{f^2}\frac{d}{dr}\left(2Uf\frac{df}{dr}\right)
&=&\frac{2}{f^2}-V-2e^{2\phi}\frac{Q^2}{f^4}.
\end{eqnarray}
For the dilaton-de Sitter black hole\cite{CZhang}, $f$ and $U$ are 
given by
\begin{eqnarray}
f=\sqrt{r(r-2D)}, U=1-\frac{2M}{r}-\frac{1}{3}\lambda r(r-2D),
\end{eqnarray}
where $\lambda$ is the cosmological constant, and $D$ is the dilaton charge.

Substituting the above expressions 
into the equations of motion, the dilaton field, dilaton charge 
and potential become
\begin{eqnarray}
e^{2\phi}&=&e^{2\phi_{0}}\left(1-\frac{2D}{r}\right),
\end{eqnarray}
\begin{eqnarray}
D&=&\frac{Q^2e^{2\phi_{0}}}{2M},
\end{eqnarray}
\begin{eqnarray}
V(\phi)&=&\frac{4}{3}\lambda+\frac{\lambda}{3}\left[e^{2(\phi-\phi_{0})}+
e^{-2(\phi-\phi_{0})}\right].
\label{dilpot}
\end{eqnarray}
where $M$ being the black hole mass, and $\phi_0$ the asymptotic constant
value of the dilaton. The dilaton potential is the sum of a constant and
two Liouville-type terms. In fact, the above structure of the dilaton potential
is essential for the existence of dilaton-de Sitter black holes\cite{poletti}.

The action for a dilaton-de Sitter black hole can hence be written as
\begin{eqnarray}
S=\int d^4X \sqrt{-g}(R-2\partial_{\mu}\phi\partial^{\mu}\phi-
\frac{4}{3}\lambda\nonumber\\
-\frac{\lambda}{3}\left[e^{2(\phi-\phi_{0})}
+ e^{-2(\phi-\phi_{0})} \right]).
\end{eqnarray}
In terms of the Hubble parameter $H$, with $H^2=\frac{\lambda}{3}$,   
the metric is given by
\begin{eqnarray}
dS^2=-\left(1-\frac{2M}{r}-r(r-2D)H^2\right)dt^2 + \nonumber\\
\left(1-\frac{2M}{r}-r(r-2D)H^2\right)^{-1}dr^2
+r(r-2D)\left(d\Omega^2\right)\label{1}
\end{eqnarray}
For $H=0$ the metric goes to GMGHS
black hole\cite{Gibbons}. For both $D=0$ and  $H=0$ it reduces to 
the well known Schwarzschild
metric. Also when $\phi=\phi_{0}=0$ the action of this space-time reduces
to the action of a Reisser-Nordstrom-de Sitter black hole.
It may be noted here that the solution  for a dilaton-de Sitter
black hole given by Eq.(\ref{1}) correspondes to a particular choice of the 
dilaton potential (\ref{dilpot}). Other solutions corresponding to
other potentials with minima $\phi_0$ at which $V(\phi_0)>0$ are possible,
but here we only consider the solution (\ref{1}) which is
written\cite{CZhang} analytically in a closed form.

\section{Particle motion in the dilaton-de Sitter metric}

The equations of motion for a test particle of mass $m$ in a curved spacetime 
with metric $g_{\mu\nu}$ are given by
\begin{eqnarray}
\frac{D^2X^{\mu}}{D\tau^2}=0,
\end{eqnarray}
where $D/D\tau$ denotes the covariant derivative with respect to proper time
$\tau$.
The equation can derived from the Lagrangian 
\begin{eqnarray}
L=\frac{1}{2} g_{\mu\nu}\dot{X}^{\mu}\dot{X}^{\nu},
\end{eqnarray}
where an overdot denotes the partial derivatives with respect to an affine
parameter $\lambda$. For consistency we chose
\begin{eqnarray}
\tau=m\lambda,
\end{eqnarray}
which is equivalent  to
\begin{eqnarray}
g_{\mu\nu}\dot{X}^{\mu}\dot{X}^{\nu}=-m^2.
\end{eqnarray}
The conjugate momenta following from the Lagrangian are given by
\begin{eqnarray}
p_{\mu}=g_{\mu\nu}\dot{X}^{\nu},
\end{eqnarray}
which calculated for the dilaton-de Sitter metric turn out to be
\begin{eqnarray}
p_{r}&=&\left[1-\frac{2M}{r} -r(r-\frac{Q^2e^{2\phi_{0}}}{M})H^2\right]^{-1}\dot{r},\nonumber\\
p_{t}&=&-\left[1-\frac{2M}{r} -r(r-\frac{Q^2e^{2\phi_{0}}}{M})H^2\right]\dot{t},
\nonumber\\
p_{\theta}&=&r(r-\frac{Q^2e^{2\phi_{0}}}{M})\dot{\theta},\nonumber\\
p_{\phi}&=&r(r-\frac{Q^2e^{2\phi_{0}}}{M})\sin^2{\theta}\dot{\phi}.
\end{eqnarray}
Since the field is isotropic, one may consider the orbit of the particle
confined to the equatorial plane i.e, $\theta=\pi/2$, and $p_{\theta}=0$.
Then the
equations of motion for the dilaton-de Sitter metric are obtained from
the variational principle to be
\begin{eqnarray}
\frac{d}{d\lambda}\left[r(r-\frac{Q^2e^{2\phi_{0}}}{M})\frac{d\phi}{d\lambda}
\right]=0\nonumber\\
\frac{d}{d\lambda}\left[\biggl(1-\frac{2M}{r}-r(r-\frac{Q^2e^{2\phi_{0}}}{M})H^2\biggr)
\frac{dt}{d\lambda}\right]=0\nonumber\\
\left[\frac{2M}{r^2}-(2r-\frac{Q^2e^{2\phi_{0}}}{M})H^2\right](\frac{dt}{d\lambda})^2\nonumber\\
-\left[1-\frac{2M}{r}-r(r-\frac{Q^2e^{2\phi_{0}}}{M})H^2\right]^{-2}\nonumber\\
\times\left[\frac{2M}{r^2}-(2r-\frac{Q^2e^{2\phi_{0}}}{M})H^2\right]
(\frac{dr}{d\lambda})^2\nonumber\\
-2\frac{d}{d\lambda}\left[(1-\frac{2M}{r}-r(r-\frac{Q^2e^{2\phi_{0}}}{M})H^2)^{-1}\frac{dr}{d\lambda}\right]\nonumber\\
+(2r-\frac{Q^2e^{2\phi_{0}}}{M})(\frac{d\phi}{d\lambda})^2
=0.
\end{eqnarray}
From the above equations we see that $p_{t}$ and $p_{\phi}$ are constants of
motion, but $p_{r}$ is not.
So we get two constants of motion corresponding to the conservation of energy 
and angular momenta denoted respectively by
\begin{eqnarray}
p_{t}&=&-E,\nonumber\\
p_{\phi}&=&\Phi.
\end{eqnarray}
In order to solve the system of equations of motion, one employs the  
Hamilton-Jacobi
method. The Hamiltonian takes the form
\begin{eqnarray}
\cal H&=&\frac{1}{2}g^{\mu\nu}p_{\mu}p_{\nu}
\end{eqnarray}
and
\begin{eqnarray}
p_{\mu}&=&\frac{\partial S}{\partial x^{\mu}}\nonumber\\
p_{\nu}&=&\frac{\partial S}{\partial x^{\nu}}
\end{eqnarray}
and $S$ is the Hamilton-Jacobi action. 

The Hamilton-Jacobi equations can be written as 
\begin{eqnarray}
-\frac{\partial S}{\partial{\lambda}}=&\cal H 
=&\frac{1}{2}g^{\mu\nu}\frac{\partial S}{\partial x^{\mu}}\frac{\partial S}
{\partial x^{\nu}} \label{2}
\end{eqnarray}
The action takes the form
\begin{eqnarray}
S&=&\frac{1}{2}m^2\lambda-Et+S_{r}+\Phi\phi,\label{3}
\end{eqnarray}
where $S_{r}$ is function of $r$.
From (\ref{2}) and (\ref{3}) one obtains
\begin{eqnarray}
&&(\frac{\partial S}{\partial r})^2
(1-\frac{2M}{r}-r(r-\frac{Q^2e^{2\phi_{0}}}{M})H^2)\nonumber\\
&&-\frac{E^2}{(1-\frac{2M}{r}-r(r-\frac{Q^2e^{2\phi_{0}}}{M})H^2)}\nonumber\\
&&+m^2\nonumber\\
&&+\frac{\Phi^2}{r(r-\frac{Q^2e^{2\phi_{0}}}{M})}=0\label{4}
\end{eqnarray}

Equation (\ref{4}) can be written in compact form:
\begin{eqnarray}
\frac{\partial S_{r}}{\partial r}&=&\sigma_{r}\sqrt{\Re}.\label{5}
\end{eqnarray}
$\sigma_{r}$ is the sign function, and $\Re$ can be written as 
\begin{eqnarray}
\Re&=&\frac{\chi}{\Delta},\label{6}
\end{eqnarray}
where
\begin{eqnarray}
\Delta&=&(1-\frac{2M}{r}-r(r-\frac{Q^2e^{2\phi_{0}}}{M})H^2)
,\label{7}
\end{eqnarray}
and 
\begin{eqnarray}
\chi=&&\frac{E^2}{(1-\frac{2M}{r}
-r(r-\frac{Q^2e^{2\phi_{0}}}{M})H^2)}\nonumber\\
&&-\frac{\Phi^2}{r(r-\frac{Q^2e^{2\phi_{0}}}{M})}\nonumber\\
&&-m^2. \label{8}
\end{eqnarray}

We can write the Hamilton-Jacobi action in terms of these functions as
\begin{eqnarray}
S=\frac{1}{2}m^2\lambda-Et+\Phi \phi+\sigma_{r}\int^r
\sqrt{\Re}dr.
\end{eqnarray}
By differentiating with respect to  $m$, $E$,  and $\Phi$ 
respectively, 
the solutions of the Hamilton-Jacobi equations can be formally written as 
\begin{eqnarray}
\lambda&=&\int_{0}^r\frac{dr}{\Delta\sqrt{\Re}},
\end{eqnarray}
\begin{eqnarray}
t&=&\int_{0}^r \frac{E dr}{\Delta\sqrt{\Re}},
\end{eqnarray}
\begin{eqnarray}
\phi&=&\int_{0}^r\frac{\Phi dr}{\Delta\sqrt{\Re}r(r-\frac{Q^2
e^{2\phi_{0}}}{M})} 
\end{eqnarray}

These equations can be expressed in the form of first-order differential
equations as
\begin{eqnarray}
\dot{r}&=&\sigma_{r}\sqrt{\Re}
\left[1-\frac{2M}{r} -r(r-\frac{Q^2e^{2\phi_{0}}}{M})H^2\right],
\end{eqnarray}
\begin{eqnarray}
\dot{t}&=&\frac{E}{\left[1-\frac{2M}{r} -r(r-\frac{Q^2e^{2\phi_{0}}}{M})H^2\right]
},
\end{eqnarray}
\begin{eqnarray}
\dot{\phi}&=&\frac{\Phi}{r(r-\frac{Q^2e^{2\phi_{0}}}{M})}.
\end{eqnarray}

A characteristic property of the four-dimensional gravitational field is the 
existence of bounded orbits located in the exterior of the black hole.
One can verify this issue in the context of dilaton-de Sitter gravity
by studying the 
circular orbits
($r=r_{0}=$ constant) that are defined by the equations
\begin{eqnarray}
\chi \Delta=0,
\end{eqnarray}
where $\Delta$ and $\chi$ are given by given by Eqs.(\ref{7}) and (\ref{8})
respectively.  
In order to obtain bound 
orbits ($E^2< m^2$) for
the metric (\ref{1}) it is also convenient to use the effective potential.

The equation of the trajectory calculated for 
the radial motion of a particle with mass $m$ in this metric (\ref{1}) is
\begin{eqnarray}
(\frac{dr}{dt})^2&=&E^2-V(r)
\end{eqnarray}
where the effective potential $V(r)$ for radial motion is given by
\begin{eqnarray}
V(r)=-\left[1-\frac{2M}{r}-r(r-\frac{Q^2e^{2\phi_{0}}}{M})H^2\right]m^2
\nonumber\\
\times \left[1+\frac{
_{{\Phi}^2}}{m^2r(r-\frac{Q^2e^{2\phi_{0}}}{M})}\right].\label{25}
\end{eqnarray}
$\Phi=r(r-\frac{Q^2e^{2\phi_{0}}}{M})\dot{\phi}$ is the $\phi$ component of the 
angular momentum.
Differentiating $V(r)$ with respect to $r$, one can find the extrema of the
effective potential corresponding to bound orbits, which are given by the
roots of the equation $dV/dr =0$ given by
\begin{eqnarray}
H^2\left[2r^6 - 10r^5D + 16r^4D^2 - 8r^3D^3\right] \nonumber\\
-2Mr^3 + 8Mr^2D + 2r^2(\Phi/m)^2 - 6Mr(\Phi/m)^2 \nonumber\\
-2rD(\Phi/m)^2 - 8MrD^2 + 8MD(\Phi/m)^2 = 0  
\label{effpot}
\end{eqnarray}
For consistency, it can be checked that for $D=0$, $H=0$, the 
above expressions return to their Schwarzschild values.
For $H=0$ one gets the effective potential for the dilaton or GMGHS black
hole. 

The motion of test particles around a charged dilaton black hole has
been studied earlier\cite{maki}. Depending on the magnitudes of the charge
and angular momentum of the test particle, the radius of the innermost stable
orbit shifts compared to that for the Schwarzschild black 
hole\cite{maki}. Here our focus is on the effects of the Hubble expansion
on the effective potential for the dilaton-de Sitter black hole. Note
first, that the magnitude of the $H$-dependent terms 
are negligible compared to the other terms in Eq.(\ref{effpot}) at small
distances from the black hole. The small value of the Hubble parameter
is obviously unable to affect the stability of inner orbits of particles
around the black hole. However, as one moves to larger distances from the black
hole, the contribution of the $H$-dependent terms increase in magnitude.
Moreover, far away from the black hole, the dilatonic contribution loses
its relevance ($D$-dependent terms become negligible). Various details of the 
geodesic structure and particle orbits in Schwarzscild-de Sitter (and 
anti de Sitter) metrics have been worked out earlier\cite{schdesit}. 
For the present case, the equation
for extrema of the effective potential ($dV/dr =0$) at large distances 
can be approximated by
\begin{eqnarray}
H^2r^5 - Mr^2 + (r-3M)(\Phi/m)^2 \approx 0
\label{effpot2}
\end{eqnarray}
It can be seen from Eq.(\ref{effpot2}) that the contribution of the Hubble
expansion towards the position of particle orbits is effective at distances
of $r \sim (M/H^2)^{1/3}$. Further analysis shows that bound orbits exist
for the dilaton-de Sitter black hole at these distance scales provided
the angular momentum per mass ($\Phi/m$) of the orbiting particle is
bounded by the relation
\begin{eqnarray}
\frac{\Phi}{m} \le \left(\frac{M^2}{H}\right)^{1/3}
\end{eqnarray} 
Beyond the above limit for ${\Phi}/m$, the condition for the existence
of minima for the effective potential (Eq.(\ref{effpot2})) is no longer
satisfied, and bound orbits for the dilaton-de Sitter black hole cease
to exist at such scales of the order of $r \sim (M/H^2)^{1/3}$ because
of the expansion of the universe.

\section{Strong field lensing by dilaton-de Sitter black holes}

The dilaton-de Sitter metric in the Schwarzschild coordinate system 
given by Eq.(\ref{1}) can be written in the form of a general spherically
symmetric and static metric useful for the analysis of gravitation lensing as
\begin{equation}
ds^2 = - A(r)dt^2  +B(r)dx^2  +C(r)\left(d\Omega^2\right)
\label{sphersym}
\end{equation}
where
\begin{eqnarray}
A(r)=\left(1-\frac{1}{r}-r^2(1-\frac{\xi}{r})H^2\right)\nonumber\\
B(r)=\left(1-\frac{1}{r}-r^2(1-\frac{\xi}{r})H^2\right)^{-1}\nonumber\\
C(r)=r^2(1-\frac{\xi}{r})
\end{eqnarray}
with $r_{s}=2M$ as the measure of distance and $\xi=\frac{Q^2e^{2\phi_{0}}}{2 M^2}$, and H is the Hubble parameter now measured in Schwarzschild radius units.
The general
formalism of strong field
gravitational lensing has been worked out by Bozza\cite{Bozza}. It is
required that the equation
\begin{equation} 
\frac{C^{\prime}}{C}=\frac{A^{\prime}}{A}
\end{equation}
admits at least one positive solution the largest of which is defined
to be the radius of the photon sphere $r_{ps}$.
For the dilaton-de Sitter metric we get  radius of photon 
sphere to be $r_{ps}=\frac{3+\xi+\eta}{4}$,
where $\eta=\sqrt{9-10\xi+\xi^2}$.

\begin{figure}[h!]
\begin{center}
\includegraphics[width=8cm]{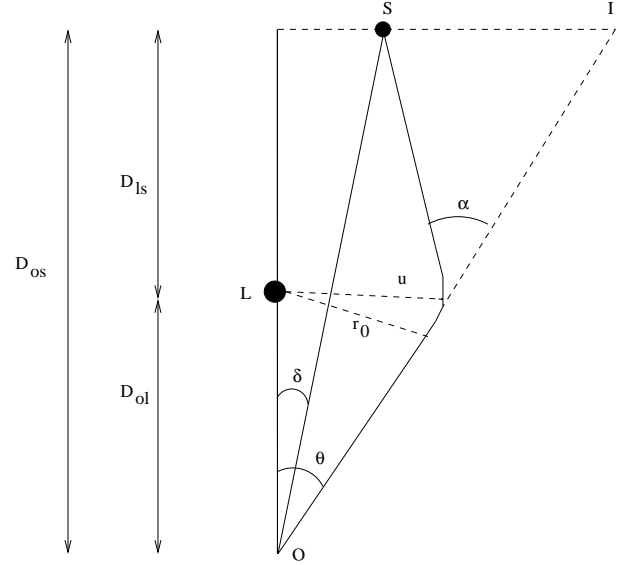}
\caption{Gravitatioanal lensing for point like mass object M. A light
ray from the source S passes the lens with an impact parameter u, and is
deflected by an angle $\alpha$. The observer sees an image I of the source at
the angular position $\theta$. }
\label{f1}
\end{center}
\end{figure}

Applications of the general framework of strong field lensing\cite{Bozza} have
been performed so far on several black hole metrics obtained from string-
and brane-inspired models\cite{Bhadra,Eiroa1,Whisker,NM2}. However, all
of the above metrics are asymptotically flat. Here we extend the application
of this formalism to the non-asymptotically flat spacetime of the 
dilaton-de Sitter metric for the first time.
A photon emanating from a
distant source and having an impact parameter $u$ will approach near
the black hole at a minimum distance $r_0$ before emerging in a different
direction (see Figure~\ref{f1}). The impact parameter is related with
the closest approach distance  by
\begin{equation}
u=\sqrt{\frac{C_{0}}{A_{0}}}
\end{equation}
where $C_{0}$ and $A_{0}$ denote the values of the functions evaluted at
$r=r_{0}$.
The impact parameter calculated at $r_{0}=r_{ps}$  is the minimum impact 
parameter which for the dilaton-de Sitter 
metric is given by
\begin{eqnarray}
u_{m}= \nonumber\\
\frac{(3+\xi+\eta)\sqrt{3+\eta-3\xi}}
{\sqrt{16(3+\xi+\eta)-64-H^2(3+\xi+\eta)^3+4\xi H^2(3+\xi+\eta)^2}}
\end{eqnarray}
It is easy to check that at $H=0$, and $\xi=0$, $u_{m}=\frac{3\sqrt{3}}{2}$, 
which is the value calculated for Schwarzschild metric, and for $H=0$, $u_{m}$
goes to GMGHS black hole value\cite{Bhadra}.

The deflection angle can be written in terms of the  distance of 
closest approach as
\begin{equation}
\alpha(r_0)=I(r_0)-\pi\label{9}
\end{equation}
where,
\begin{eqnarray}
I(r_{0})=\int_{r_{0}}^\infty 
\frac{2\sqrt{B}}{\sqrt{C}\sqrt{\frac{C}{C_{0}}\frac{A_{0}}{A}-1}} dr.\label{10}
\end{eqnarray}
Substituting the values of the various quantities for the dilaton-de Sitter
metric, we get
\begin{eqnarray}
I(r_{0})=2\int_{r_{0}}^\infty 
\frac{dr}{r\sqrt{F(r,r_{0})}}
\end{eqnarray}
where
\begin{equation}
F(r,r_{0})=
\end{equation}
\begin{eqnarray}
(\frac{r}{r_{0}})^2(1-\frac{\xi}{r})^2
(1-\frac{\xi}{r_0})^{-1}(1-\frac{1}{r_{0}})
-(1-\frac{\xi}{r})(1-\frac{1}{r}))\nonumber
\label{12}
\end{eqnarray}
Puttting $z=1-\frac{r_{0}}{r}$,
one gets
\begin{equation}
I(r_{0})=\int_0^1 R(z,r_{0})f(z,r_{0}) dz.\label{13}
\end{equation}
where
\begin{equation}
R=2\frac{\sqrt{(1-\frac{\xi}{r_{0}})}}{1-\frac{\xi}{r_{0}}+z\frac{\xi}{r_{0}}},
\end{equation}
is regular for all values of $z$,
and
\begin{eqnarray}
f(z,r_{0})=[(1-\frac{1}{r_{0}}-(1-\frac{1-z}{r_{0}})\nonumber\\
\times(1-z)^2(1-\frac{\xi}{r_{0}})(1-\frac{1-z}{r_{0}}\xi)^{-1}]^{-1/2}
\end{eqnarray}
 diverges for $z \rightarrow 0$. We expand the argument of the squareroot 
in $f(z,r_{0})$ to the second order in $z$ and get,
\begin{eqnarray}
f({z,r_{0}})\sim f_{0}(z,r_{0})&=&\frac{1}{\sqrt{\alpha z + \beta z^2}}
\end{eqnarray}
with
\begin{eqnarray}
\alpha &=& -\frac{3}{r_{0}}+2\nonumber\\
&&+\frac{\xi}{r_{0}}(1-\frac{\xi}
{r_{0}})^{-1}(1-\frac{1}{r_{0}})\label{14}
\end{eqnarray}
and
\begin{eqnarray}
\beta=\frac{1}{2}[\frac{6}{r_{0}}-2+\frac{\xi}{r_{0}}(1-\frac{\xi}{r_{0}})^{-1}
(\frac{6}{r_{0}}-4)\nonumber\\
-2(\frac{\xi}{r_{0}})^2
(1-\frac{\xi}{r_{0}})^{-2}(1-\frac{1}{r_{0}})]\label{15}
\end{eqnarray}
Again, $\alpha$ and $\beta$ reduce to their Schwarzschild values\cite{Bozza}
for $\xi=0$, and they are the same for the
GMGHS metric\cite{Bhadra} since the $H$-dependence has already dropped out.

\vskip 0.2in

The  total deflection angle is given by $I(r_{0})=I_{D}(r_{0})+I_{R}(r_{0})$,
where
\begin{equation}
I_{D}(r_{0})=\int_0^1 R(0,r_{ps})f_{0}(z,r_{0}) dz,
\end{equation}
 includes the divergent part with $R(0,r_{ps})=\frac{2}{\sqrt{1-\frac{\xi}{r_{ps}}}}$.
The second term in the deflection angle given by
\begin{eqnarray}
&I_{R}(r_{0})=\int_0^1 g(z,r_{0})dz\nonumber\\
&=\int_0^1[R(z,r_{0})f(z,r_{0})-R(0,r_{ps})f_{0}(z,r_{0})]dz
\end{eqnarray}
is the original integral with the divergence substracted.
One can expand $I_{R}(r_{0})$ in power of $(r_{0}-r_{ps})$ and considering the
first expansion term we get
\begin{eqnarray}
I_{R}(r_{0})&=&\int_0^1 g(z,r_{ps})dz+ O(r_{0}- r_{ps})\label{16}
\end{eqnarray}
This integral can be evaluated and the regular term of the deflection angle for the dilaton-de Sitter metric is obtained to be
\begin{eqnarray}
I_{R}(r_{ps})=\frac{2}{\sqrt{1-\frac{\xi}{r_{ps}}}\sqrt{\beta}}[
ln(4\sqrt{\beta})\nonumber
\end{eqnarray}
\begin{equation}
-ln(\frac{2\beta+\gamma}{\sqrt{\beta}}
+2\sqrt{\beta+\gamma+\delta})]\label{17}
\end{equation}
where the expressions for $\gamma$ and $\delta$ are given by,
\begin{eqnarray}
\gamma=\frac{r_{ps}-4\xi+\xi r_{ps}}{r_{ps}(\xi-r_{ps})}
\label{18}
\end{eqnarray}

\begin{eqnarray}
\delta=\frac{\xi}{ r_{ps}(\xi-r_{ps})}
\label{19}
\end{eqnarray}
All the coefficients are evaluated at the point $r_{0}=r_{ps}$, and 
we get only the regular part of the deflection angle\cite{Bozza}.
Setting $\xi=0$ one obtains all the coefficients for the Schwarzschild
metric.

The expression for the strong field limit of the deflection angle may
be written as\cite{Bozza}
\begin{eqnarray}
\alpha_{\theta}=-\bar{a}ln[\frac{\theta D_{ol}}{u_{rps}}-1]+\bar{b}, \label{20}
\end{eqnarray}
where, 
\begin{eqnarray}
\bar{a}&=&\frac{R(0,r_{ps})}{2\sqrt{\beta_{rps}}}\label{21}
\end{eqnarray}
\begin{eqnarray}
R(0,r_{ps})&=&2\frac{\sqrt{3+\eta+\xi}}{\sqrt{3+\eta-3\xi}}
\end{eqnarray}
\begin{eqnarray}
\bar{b}&=&-\pi+b_{R}+\bar{a}ln\left[\frac{2\beta_{r_{ps}}}
{1-\frac{1}{r_{ps}}-r_{ps}^2(1-\frac{\xi}{r_{ps}})H^2}
\right]\nonumber\\\label{22}
\end{eqnarray}
\begin{eqnarray}
b_{R}=\frac{R(0,r_{ps})}{2\sqrt{\beta_{r_{ps}}}}[2ln(4\sqrt
{\beta_{r_{ps}}})\nonumber\\
-ln(\frac{2\beta_{r_{ps}}+\gamma_{r_{ps}}}{\sqrt{\beta_{r_{ps}}}}
+2\sqrt{\beta_{r_{ps}}+\gamma_{r_{ps}}+\delta_{r_{ps}}})]
\end{eqnarray}
Note that the expression for $\bar{b}$ given in Eq.(\ref{22}) contains
a $H$-dependent term which introduces a small correction to the lensing
angle. 

In strong gravitational lensing there may exist
$n$ relativistic images given by the number of times a light ray loops
around the black hole. The positions of the source and the images are 
related through the lense equation derived by Virbhadra and Ellis\cite{Fritte}
given by
\begin{equation}
{\mathrm tan} \delta = {\mathrm tan} \theta - \frac{D_{ds}}{D_s}
[{\mathrm tan} \theta + {\mathrm tan}(\alpha - \theta)]
\label{lenseq}
\end{equation}
The magnification $\mu_n$ of the $n$-th relativistic image is given
by\cite{Bozza}
\begin{equation}
\mu_n = \frac{1}{(\delta/\theta)\partial \delta \partial \theta}
\vert_{\theta_n} \simeq \frac{u_m^2 e_n (1+e_n)D_s}{\overline{a}\delta
D_{ds}D_d^2}
\label{magnif}
\end{equation}
where
$e_n = {\mathrm e}^{(\overline{b} -2n\pi)/\overline{a}}$.
The expressions for the various lensing observables can be obtained
in terms of the metric parameters.
For $n \to \infty$ an observable $\theta_{\infty}$
can be defined\cite{Bozza} representing the asymptotic position approached
by a set of images.
The minimum impact parameter can then be  obtained as
\begin{equation}
u_m=D_{d} \theta_{\infty}
\label{thetainfty}
\end{equation}
In the simplest situation where only the
outermost image $\theta_1$ is resolved as a single image, while
all the remaining ones are packed together at $\theta_\infty$,
two lensing observables can be defined as\cite{Bozza}
\begin{eqnarray}
{\cal S}=\theta_1-\theta_\infty
\end{eqnarray}
representing the separation between the first
image and the others, and
\begin{eqnarray}
{\cal R}=\frac{\mu_1}{\sum\limits_{n=2}^\infty \mu_n}
\end{eqnarray}
corresponding to the ratio between the flux of the first
image and the flux coming from all the other images.

In terms of the deflection angle parameters
$\overline{a}$ and $\overline{b}$, these observables can be written
as\cite{Bozza}
\begin{equation}
{\cal S}= \theta_\infty e^{\overline{b}/\overline{a}
- 2\pi/\overline{a}}
\label{obs-s}
\end{equation}
\begin{equation}
{\cal R}=e^{2\pi/\overline{a}}
\label{obs-r}
\end{equation}
The above equations (\ref{obs-s}) and (\ref{obs-r}) can be inverted to express
$\overline{a}$ and $\overline{b}$ in terms of the image separation ${\cal S}$
and the flux ratio ${\cal R}$. Therefore the knowledge of these two observables
can be used to reconstruct the deflection angle given by Eq.(\ref{20}).
The aim of strong field gravitational lensing is to detect the
relativistic images corresponding to specific lensing candidates and
measure their separations and flux ratios. Once
this is accomplished, the observed data could be compared with the
theoretical coefficients obtained using various metrics. A precise set
of observational data for strong gravitational lensing, if obtained,  could
therefore be able discriminate between different models of gravity.

Lensing observables such as the angular position of the relativistic images
($\theta_{\infty})$, the angular separation of the outermost relativistic 
image with the remaining bunch of relativistic images (${\cal S}$), and the 
relativistic magnification of the outermost relativistic image with 
respect to the other relativistic images ($r_m$) have been computed earlier
for the galactic centre black hole considering various candidate geometries, 
viz. the Schwarzschild black
hole\cite{Bozza}, the braneworld black hole with negative tidal 
charge\cite{dadhich,Whisker,NM2},
and the Reissner-Nordstrom black hole (which can also be considered as 
a GMGHS black 
hole\cite{Gibbons,Bhadra} with dilaton potential minimum $\phi_{0}=0$). 
All these metrics have aroused interest recently as black hole candidates 
representing gravity modified in separate contexts.
It should be mentioned here
that according to recent observational results\cite{bhspin}, 
supermassive black holes including the one in our
galactic centre seem to be spinning at considerable rates as described by
the Kerr solution, but there is lack of evidence regarding any charged 
astrophysical black holes. From the theoretical viewpoint, no black hole 
solution has yet been found
which incorporates both the dilaton and de Sitter expansion for the
Kerr black hole, though there exit some interesting applications of particle
motion in the Kerr-de Sitter metric\cite{kerrdesit}.  Further, 
strong gravitational
lensing by the Kerr metric itself is endowed with additional features, and
the full phenomenological implications are in the process of
being worked out\cite{bozza2}.

Considering the supermassive black hole in the galactic center 
(with its mass $4.3\times10^{6}M_{\bigodot}$ and its distance from us 
$8 kpc$\cite{recent}) as a 
charged dilaton 
black hole in a cosmological background, one can estimate the observables
of strong lensing. 
However, due to the small value of the Hubble 
parameter ($H = 7.7\times10^{-27}$s) 
the de Sitter expansion makes a rather negligible contribution to the actual
values of the lensing observables. This is apparent from considering the
minimum impact parameter given by Eq.(46), which can be expanded in terms of
powers of $H$ as
\begin{eqnarray}
u_m &=& \bigl(u_m\bigr)_{\mathrm GMGHS}\left(1 + \frac{(3 + \xi + \eta)(3-3\xi+\eta)^{1/2}{H^2}}{2[16(3+\xi+\eta)-64]}\right)\nonumber\\
 &+& O(H^4)
\end{eqnarray}
where $\bigl(u_m\bigr)_{\mathrm GMGHS}$ denotes the minimum impact parameter
for the GMGHS metric. Hence one sees that the Hubble expansion
makes a rather tiny modification ($O(H^2)$) to the observable values, that are 
too insignificant to enable, for example, the dilaton-de Sitter metric to 
be distinguished
from the GMGHS metric.

\section{Conclusions}

In this paper we have discussed several features of the motion of particles 
and light in the 
spacetime of a charged dilaton black hole in a cosmological background.
The black hole solution recently
obtained\cite{CZhang} corresponding to the dilaton-de Sitter metric 
could be relevant for the study of string theoretic
implications in the context of the presently accelerating universe. 
The conserved quantities in the dilaton-de Sitter metric have been identified
and the Hamilton-Jacobi method employed for the equations of motion.
We have also obtained the effective potential for this metric in order to
analyse the existence of bound orbits in this metric in an expanding
universe. Close to the black hole the dilatonic contribution in the 
effective potential dominates
over that of the Hubble term which is rather negligible. However, we find that
at large distances from the black hole the effective potential could be
significantly affected by the Hubble expansion. We have derived an
expression for the upper value of the angular momentum per mass of
the orbiting particle up to which bound orbits could exist in this metric.

We have further investigated strong field gravitational lensing in the
dilaton-de Sitter metric. 
Gravitational lensing promises to be a powerfull tool in future observations
for probing the strong field character of gravity 
that may be able to incorporate signatures of extra
dimensions as in string or brane models. Here we have extended the application
of  the standard strong lensing framework\cite{Bozza} used previously for only
asymptotically flat spacetimes, 
for obtaining the expressions for the different
lensing variables as functions of the metric parameters of the 
non-asymptotically flat dilaton-de Sitter metric. 
We note however, that the  value of the Hubble
parameter is too small to impact the position of relativistic images and
other strong lensing quantities in any observationally significant way.
We conclude with the rider that though our present analysis
may not provide accurate estimates of the actual position of stable orbits
or of lensing observables
for any astrophysical black hole in the expanding universe, these calculations
may serve to motivate
future analysis using more advanced theoretical techniques suitable to handle
the realistic situations including effects such as black hole rotation.

\end{document}